\documentclass[twocolumn,showpacs,preprintnumbers,amsmath,amssymb,superscriptaddress]{revtex4}

\usepackage{graphicx}
\usepackage{dcolumn}
\usepackage{bm}
\usepackage{mathrsfs}

\newcommand{\norm}[1]{\lVert#1\rVert}

\makeatletter
\def\ExtendSymbol#1#2#3#4#5{\ext@arrow 0099{\arrowfill@#1#2#3}{#4}{#5}}
\def\RightExtendSymbol#1#2#3#4#5{\ext@arrow 0359{\arrowfill@#1#2#3}{#4}{#5}}
\def\LeftExtendSymbol#1#2#3#4#5{\ext@arrow 6095{\arrowfill@#1#2#3}{#4}{#5}}
\makeatother

\newcommand\myArrow[2][]{\ExtendSymbol{\Leftarrow}{=}{\Rightarrow}{#1}{#2}}

\begin{document}

\title{General Greenberger-Horne-Zeilinger theorem of cluster states}

\author{Li Tang}
 \email{litang@mail.ustc.edu.cn}
\affiliation{Hefei National Laboratory for Physical Sciences at
Microscale and Department of Modern Physics, University of Science
and Technology of China, Hefei, Anhui 230026, China}

\author{Zeqian Chen}
\affiliation{State Key Laboratory of Magnetic Resonance and Atomic
and Molecular Physics and United\\ Laboratory of Mathematical
Physics, Wuhan Institute of Physics and Mathematics, Chinese Academy
of Sciences, 30 West District, Xiao-Hong-Shan, P.O.Box 71010, Wuhan
430071, China}

\author{Zeng-Bing Chen}
\affiliation{Hefei National
Laboratory for Physical Sciences at Microscale and Department of
Modern Physics, University of Science and Technology of China,
Hefei, Anhui 230026, China}

\date{December 2008}%

\begin{abstract}
In this paper, we show that there are eight distinct forms of the
Greenberger-Horne-Zeilinger (GHZ) argument for the four-qubit
cluster state $|\phi_4\rangle$ and forty eight distinct forms for
the five-qubit cluster state $|\phi_5\rangle$ in the case of the
one-dimensional lattice. The proof is obtained by regarding the pair
qubits as a single object and constructing the new Pauli-like
operators. The method can be easily extended to the case of the
N-qubit system and the associated Bell inequalities are also
discussed. Consequently, we present a complete construction of the
GHZ theorem for the cluster states of N-qubit in the case of the
one-dimensional lattice.
\end{abstract}

\pacs{03.65.Ud, 03.67.Mn, 03.67.Pp}
\maketitle

\section{introduction}

Bell's inequality \cite{Bell} indicates that certain statistical
correlations predicted by quantum mechanics for measurements on
two-qubit ensembles cannot be understood within a realistic picture
based on Einstein, Podolsky, and Rosen's (EPR's) notion of local
realism \cite{EPR}. However, there is an unsatisfactory feature in
the derivation of Bell's inequality that such a local realistic and,
consequently, classical picture can explain perfect correlations and
is only in conflict with statistical prediction of quantum
mechanics. Strikingly enough, the Greenberger-Horne-Zeilinger
(GHZ's) theorem exhibits that the contradiction between quantum
mechanics and local realistic theories arises even for definite
predictions on a four-qubit system \cite{GHZ}. Mermin \cite{Mermin}
subsequently refined the original GHZ argument on a three-qubit
system. Motivated by this discovery, a growing amount of interest in
studying the various types of the GHZ argument has been apparent in
recent literatures \cite{GHZth}. In particular, the GHZ argument for
cluster states is presented in \cite{SASA}. In this paper, we will
present a complete construction of the GHZ argument for the cluster
states of N-qubit in the case of the one-dimensional lattice.

As is well known, the cluster states are suitable multi-qubit states
for universal, scalable quantum computation \cite{RBB} and can be
used in the quantum error-correction code \cite{GSW}. There have
been a number of experimental systems proposed as candidates for the
generation of cluster states, and several of them have been
implemented, e.g., creation of six-photon cluster states with
verifiable six-partite entanglement by a linear optical elements
\cite{LP}.

Following \cite{RBB} we give a brief review of the definition and
main properties of the cluster states in the case of an
one-dimensional lattice with an open segment, which we denote
$|\phi_N \rangle.$ The cluster state $|\phi_N \rangle$  is
determined by the set of eigenvalue equations
\begin{align}
E_a|\phi_N\rangle=|\phi_N\rangle,
\end{align}
with the correlations  operators
\begin{align}
E_a=X_a\bigotimes_{b\in \textrm{neigh}(a)}Z_b,
\end{align}
where neigh$(a)$ is the set of all neighbors of $a,$ $X=\sigma_x,$
$Y=\sigma_y,$ and $Z=\sigma_z.$ For such a lattice, $| \phi_2
\rangle$ and $| \phi_3 \rangle$ are locally equivalent to a
maximally entangled Bell state and  a GHZ state, respectively.
Indeed,
\begin{align}
|\phi_3 \rangle = \frac{1}{\sqrt{2}} \big ( | + \rangle | 0 \rangle
| + \rangle + |- \rangle | 1 \rangle |- \rangle \big ).
\end{align}
Hereafter, the one-qubit states are defined as usual as $Z | 0
\rangle = | 0 \rangle,$ $Z | 1 \rangle = - | 1 \rangle,$ and $X |
\pm \rangle = \pm | \pm \rangle.$ Moreover, the four-qubit cluster
state  is
\begin{equation}\begin{split}
|\phi_4\rangle = & \frac{1}{2} \Big (
|+\rangle|0\rangle|+\rangle|0\rangle
+|+\rangle|0\rangle|-\rangle|1\rangle\\
&~~ + |-\rangle|1\rangle|-\rangle|0\rangle
+|-\rangle|1\rangle|+\rangle|1\rangle \Big ).
\end{split}\end{equation}
Note that the state $| \phi_N \rangle$ is not locally equivalent to
the N-qubit GHZ state whenever $N \geq 4$. In particular  in the
case of  $|\phi_4\rangle$,   a Bell inequality based on the GHZ
argument can be derived \cite{SASA, Cabello} and has been tested in
laboratory \cite{WKV}. It acts as a strong entanglement witness,
takes the quantum upper bound 4 as its maximum, while the four-qubit
GHZ state does not violate it at all. However no complete
construction of the GHZ theorem   has been reported for the cluster
states of N-qubit in the case of the one-dimensional lattice.

This paper is organized as follows. In Sec.~\ref{sec:3 qubits}, we
give a short recall of the general GHZ theorem for the three-qubit
cluster state $|\phi_3\rangle$ (as the three-qubit GHZ state
$|\mathrm{GHZ}_3\rangle$). In Sec.~\ref{sec:4 and 5 qubits}  we show
how our method  is applied to construct the GHZ argument for
$|\phi_4\rangle$ and $|\phi_5\rangle,$ respectively. The proof is
obtained by regarding the pair qubits as a single object and
constructing the new Pauli-like operators. In Sec.~\ref{sec:N
qubits} the  method is generalized to the case of the N-qubit
system. In Sec.~\ref{sec:Bell} we discuss the Bell inequalities
associated with the GHZ argument for the cluster states of N-qubit.
A final section, Sec.~\ref{sec:summary}, summarizes our conclusion.

\section{General GHZ theorem for $|\phi_3 \rangle$}
\label{sec:3 qubits}

The three-qubit cluster state $|\phi_3\rangle$ is defined by
\begin{align}
X_1 Z_2 I_3 |\phi_3\rangle & =1, &(E_1) \\
Z_1 X_2  Z_3|\phi_3\rangle & =1,&(E_2)\\
I_1  Z_2 X_3|\phi_3\rangle & =1. &(E_3)
\end{align}
Let us recall that the scenario for the GHZ argument of the cluster
state $|\phi_3 \rangle$ is the following: Particles 1, 2, 3 move
away from each other. At a given time, an observer, Alice, has
access to particle 1, a second observer, Bob, has access to particle
2 and  a third observer, Charlie, has access to particle 3. The GHZ
theorem is obtained by involving the algebra of Pauli matrices
\begin{align}
~~~~&Z_1 X_2 Z_3|\phi_3\rangle  = 1,  &(E_2)\\
&Y_1 Y_2 Z_3|\phi_3\rangle  = 1,   &(E_1\times E_2)\\
&Y_1 X_2 Y_3|\phi_3\rangle  =-1,& -(E_1\times E_2 \times E_3)\\
&Z_1 Y_2 Y_3|\phi_3\rangle  =1.   &(E_2\times E_3)
\end{align}
According to EPR's criterion of local realism, Eqs.~(8)-(11) allow
three observers Alice, Bob and Charlie to predict the following
relation between the values of the elements of reality
\begin{align}
z_1x_2z_3 & = 1,\hspace{3cm}\\
y_1y_2z_3 & = 1,\\
y_1x_2y_3 & = -1,\\
z_1y_2y_3 & = 1.
\end{align}
However, Eqs.~(12)-(15) are inconsistent, because when we take the
product of Eqs.~(12)-(15), the value of the left-hand side is $1$,
while the right-hand side is $-1$. This suggests that Eqs.~(8)-(11)
exactly exhibit an ``all versus nothing'' contradiction between
quantum mechanics and EPR's local realism. Here  $z_1x_2z_3$ etc.\
are shorthand for the values of the elements of reality
$\nu(Z_1)\nu(X_2)\nu(Z_3)$.

It is proved in \cite{Chen} that not only is the GHZ argument of
three qubits valid merely for $ |\phi_3\rangle$ but also there is
only one form of the GHZ argument for $ |\phi_3\rangle$ under
locally unitary equivalences, namely,
\begin{align}
\{Z_1X_2Z_3, Y_1Y_2Z_3,Y_1X_2Y_3, Z_1Y_2Y_3\}.
\end{align}
The associated Bell operator reads as
\begin{align}
\mathcal {B}_3=&Z_1X_2Z_3+Y_1Y_2Z_3-Y_1X_2Y_3+Z_1Y_2Y_3 \nonumber \\
=&(1+E_1)E_2(1+E_3),
\end{align}
which reaches 4 when evaluated on $|\phi_3\rangle$, and no other
state can ever give a larger value.

Contrary to the three-qubit case, there are distinct forms of the
GHZ argument for the GHZ state of four-qubit. In \cite{TC} the
authors present a complete construction of the GHZ argument for the
four-qubit GHZ state, of which there are nine distinct forms. As
shown in \cite{SASA}, the GHZ argument is also valid for the cluster
state $| \phi_4 \rangle$ except for the GHZ state $| \mathrm{GHZ}_4
\rangle.$ Since each form of the GHZ argument admits a Bell
inequality, it is meaningful to give a complete construction of the
GHZ argument for $| \phi_4 \rangle$£¬ which will be done in the next
section.

\section{ General GHZ theorem for $|\phi_4\rangle$ and $|\phi_5\rangle$}
\label{sec:4 and 5 qubits}

The four-qubit cluster state $|\phi_4\rangle$ is defined by
\begin{align}
&X_1 Z_2 I_3I_4 |\phi_4\rangle=1,  &(E_1)\\
&Z_1 X_2  Z_3I_4|\phi_4\rangle=1, &(E_2)\\
&I_1  Z_2 X_3 Z_4|\phi_4\rangle=1, &(E_3)\\
&I_1I_2 Z_3 X_4|\phi_4\rangle=1.  &(E_4)
\end{align}
Specifically, $|\phi_4 \rangle$  can be projected into  a variation
of $|\phi_3\rangle$ as
\begin{eqnarray}
|\phi_4\rangle=\frac{1}{\sqrt{2}} \Big (
|+\rangle_1|0\rangle_2|+\rangle'_{3,4}+|-\rangle_1|1\rangle_2|-\rangle'_{3,4}
\Big ),
\end{eqnarray}
with two Bell basis vectors defined on $\mathbb{C}^2\otimes
\mathbb{C}^2$
\begin{eqnarray}
\begin{split}
|+\rangle'_{i,j} = & \frac{1}{\sqrt{2}}(|0\rangle_i|+\rangle_j+|1\rangle_i|-\rangle_j),\\
|-\rangle'_{i,j} = &
\frac{1}{\sqrt{2}}(|0\rangle_i|+\rangle_j-|1\rangle_i|-\rangle_j).
\end{split}
\end{eqnarray}
If we regard the pair qubit 3 and 4 as a single object, it makes
sense to discuss the GHZ argument as the one in Sec.~\ref{sec:3
qubits}
 since
$|+\rangle'_{3,4}$ and $|-\rangle'_{3,4}$ could be treated as an
orthonormal basis in $\mathbb{C}^2.$ This motivates us to introduce
four groups of Pauli-like operators as
\begin{align}
\begin{split}
Y'_{3,4} & \in \{-X_3Y_4,    Y_3Z_4\},\\
Z'_{3,4} & \in \{I_3X_4,   Z_3I_4\},
\end{split}
\end{align}
with
\begin{align}
X_{3,4}'=-iY'_{3,4}Z'_{3,4} \in \{X_3Z_4, Y_3Y_4\}, \end{align}
 since
$X'_{3,4},Y'_{3,4},Z'_{3,4}$ satisfy the usual algebraic identities
of Pauli's matrices \cite{Pauli} (see more details in
Table~\ref{tab:34}):
\begin{align}
\begin{split}
[X'_{3,4},Y'_{3,4}] & = 2iZ'_{3,4},\\
[Y'_{3,4},Z'_{3,4}] & = 2iX'_{3,4},\\
[Z'_{3,4},X'_{3,4}] & = 2iY'_{3,4}.
\end{split}
\end{align}

\begin{table}
\caption{\label{tab:34}$Y'_{3,4}$, $Z'_{3,4}$, $X'_{3,4}$.}
\begin{ruledtabular}
\begin{tabular}{rcl}
$ Y'_{3,4}$& $Z'_{3,4}$ & $X'_{3,4}$   \\
\hline
$-X_3Y_4$ & $I_3X_4$ &$X_3Z_4$\\
$-X_3Y_4$ & $Z_3I_4$ &$Y_3Y_4$\\
$Y_3Z_4$ & $I_3X_4$ &$Y_3Y_4$\\
$Y_3Z_4$  & $Z_3I_4$ &$X_3Z_4$\\
\end{tabular}
\end{ruledtabular}
\end{table}

Choosing $X'_{3,4}$ representation
$\{|+\rangle'_{3,4},|-\rangle'_{3,4}\},$ one has
\begin{align}
\begin{split}
X'_{3,4} | \pm \rangle'_{3,4} & = \pm | \pm \rangle'_{3,4},\\
Y'_{3,4} | \pm \rangle'_{3,4} & = \mp i | \mp \rangle'_{3,4},\\
Z'_{3,4} | \pm \rangle'_{3,4} & = | \mp \rangle'_{3,4},\\
I'_{3,4} | \pm \rangle'_{3,4} & = |\pm\rangle'_{3,4},
\end{split}
\end{align}
with $I'_{3,4}\in\{Z_3X_4, I_3I_4\}.$ Apparently, according to the
unique form Eq.~(16) of the GHZ argument for $ |\phi_3\rangle$ in
Sec.~\ref{sec:3 qubits},   the GHZ argument for $|\phi_4\rangle$
(written as Eq.~(22)) can be expressed in terms of Pauli-like
operators as
\begin{align}
\{Z_1X_2Z'_{3,4}, Y_1Y_2Z'_{3,4},Y_1X_2Y'_{3,4}, Z_1Y_2Y'_{3,4}\}.
\end{align}
Combining  Eq.~(24) and  Eq.~(28), we obtain the four distinct forms
of the GHZ argument for $ |\phi_4\rangle$ as follows:
\begin{align}
&\{ZXIX,~YYIX,~YXXY,~ZYXY\},\\
&\{ZXIX,~YYIX,~YXYZ,~ZYYZ \},\\
&\{ZXZI,~YYZI,~YXXY,~ZYXY \},\\
&\{ZXZI,~YYZI,~YXYZ,~ZYYZ\}.
\end{align}
For simplicity, we denote $ZXIX$ etc.\ are shortcuts for
$Z_1X_2I_3X_4$.

Furthermore, we can rewrite $|\phi_4\rangle$ in the following
variation of $|\phi_3\rangle$
\begin{align}
|\phi_4\rangle=\frac{1}{\sqrt{2}}
(|+\rangle''_{1,2}|0\rangle_3|+\rangle_4+|-\rangle''_{1,2}|1\rangle_3|-\rangle_4),
\end{align}
with
\begin{align}
\begin{split}
&|+\rangle''_{i,j}=\frac{1}{\sqrt{2}}(|+\rangle_i|0\rangle_j+|-\rangle_i|1\rangle_j),\\
&|-\rangle''_{i,j}=\frac{1}{\sqrt{2}}(|+\rangle_i|0\rangle_j-|-\rangle_i|1\rangle_j).
\end{split}
\end{align}
Similarly, by regarding the pair qubit 1 and 2 as a single object
and introducing another  four groups of operators:
\begin{align}
\begin{split}
Y''_{1,2} & \in \{Z_1Y_2,  -Y_1X_2\},\\
Z''_{1,2} & \in \{X_1I_2, I_1Z_2\},
\end{split}
\end{align}
with
\begin{align}
X''_{1,2} = -iY''_{1,2}Z''_{1,2}  \in \{Z_1X_2, Y_1Y_2\},
\end{align}
 which also satisfy the algebraic identities Eq.~(26), we have
the following four forms of the GHZ argument for $ |\phi_4\rangle$
\begin{align}
&\{XIXZ,~ZYYZ,~ZYXY,~XIYY\},\\
&\{XIXZ,~YXYZ,~YXXY,~XIYY \},\\
&\{IZXZ,~ZYYZ,~ZYXY,~IZYY \},\\
&\{IZXZ,~YXYZ,~YXXY,~IZYY\}.
\end{align}

\begin{figure}
\includegraphics[width=0.35 \textwidth,keepaspectratio]{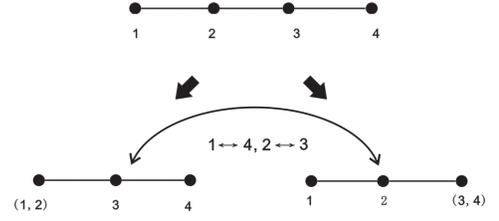}
\caption{\label{fig:4 qubits}Diagram of the two ways to derive the
GHZ argument for $|\phi_4\rangle$ and  they can be transformed to
each other by simultaneously permuting qubit 1 and 4, qubit 2 and 3
based on the basic symmetry of $|\phi_4\rangle$.}
\end{figure}

We would like to point out that the form Eq.~(37) is just the one of
the GHZ argument presented in \cite{SASA}. Moreover, Eqs.~(37)-(40)
can be obtained from Eqs.~(29)-(32) by simultaneously permuting qubit
1 and 4, qubit 2 and 3 based on the basic symmetry of
$|\phi_4\rangle$ (as shown in Fig.~\ref{fig:4 qubits}).  Since each
form of the GHZ argument for the four-qubit system  of the
one-dimensional lattice can be reduced to the three-qubit case, our
method is universal to find all forms of the GHZ argument for the
four-qubit system from the three-qubit case. Hence for the
four-qubit cluster state $|\phi_4 \rangle,$ we have obtained all the
eight distinct forms of the GHZ argument.

On the other hand, our method suggests that there are some states
other than both $|\phi_4 \rangle$ and the GHZ state, which also
exhibit the GHZ argument, for instant,
\begin{align*}
\begin{split}
|\Phi \rangle' =& \alpha (|+0+0\rangle+|-1+1\rangle)\\
&~~ + \beta(|+0-1\rangle+|-1-0\rangle)
\end{split}
\end{align*}
($|\alpha|^2+|\beta|^2 = 1/2$) are the common eigenstates of the
observable set Eq.~(29) on which those observables assume values
that refute the attempt to assign values only required to have them
by EPR's local realism.

In general, an N-qubit cluster state is given by \cite{RBB}
\begin{align}
|\phi_N\rangle=\frac{1}{2^{N/2}}\bigotimes_{a=1}^N(|0\rangle_a+|1\rangle_a\sigma_z^{a+1}),
\end{align}
with $\sigma_z^{N+1}=1$. Expanding the last two terms  leads to
\begin{align}
|\phi_N\rangle
&=\frac{1}{2^{(N-1)/2}}\bigotimes_{a=1}^{N-1}(|0\rangle_a+|1\rangle_a\sigma_z^{a+1})|+\rangle_N\nonumber\\
&=\frac{1}{2^{(N-2)/2}}\bigotimes_{a=1}^{N-2}(|0\rangle_a+|1\rangle_a\sigma_z^{a+1})|+\rangle_{N-1,N}'\nonumber\\
 &\Longrightarrow |\phi_{N-1}\rangle. ~~~~\text{(as a variation of
 $|\phi_{N-1}\rangle$)}
\end{align}

On the other hand, we may reformulate $|\phi_N\rangle$ as
\begin{eqnarray}
|\phi_N\rangle=\frac{1}{2^{N/2}}\bigotimes_{a=N}^1(|0\rangle_a+|1\rangle_a\sigma_z^{a-1}).
\end{eqnarray}
Expanding the first two terms in Eq.~(43)  leads to
\begin{align}
|\phi_N\rangle
&=\frac{1}{\sqrt{2}}(|0\rangle_N+|1\rangle_N\sigma_z^{N-1})|\phi_{N-1}\rangle\nonumber\\
&=\frac{1}{\sqrt{2}}(|+\rangle_N|0\rangle_{N-1}+
|-\rangle_N|1\rangle_{N-1}\sigma_z^{N-2})|\phi_{N-2}\rangle
\nonumber\\
&=\frac{1}{\sqrt{2}}(|0\rangle''_{N,N-1}+|1\rangle''_{N,N-1}\sigma_z^{N-2})|\phi_{N-2}\rangle\nonumber\\
&\Longrightarrow|\phi_{N-1}\rangle. ~~~~\text{(as a variation of
 $|\phi_{N-1}\rangle$)}
\end{align}

From Eqs.~(41-44), we conclude that the N-qubit cluster state
$|\phi_N\rangle$ can be projected into two variations of
$|\phi_{N-1}\rangle$ by  making proper combination of the $N$ qubits
in the above two ways.

For the purpose of further discussion, it is convenient to define
the following class of multiqubit Pauli-like operators in the
following recursive way:
\begin{align}
\begin{split}
Y'_{k,N} & \in \{-X_kY_{k+1,N}', Y_kZ_{k+1,N}'\},\\
Z'_{k,N} & \in  \{I_kX_{k+1,N}', Z_kI_{k+1,N}'\},\\
I'_{k,N} & \in  \{Z_kX_{k+1,N}', I_kI_{k+1,N}'\},
\end{split}
\end{align}
with $X'_{k,N} = -iY'_{k,N}Z'_{k,N} \in \{X_kZ_{k+1,N}',
Y_kY_{k+1,N}'\}$, and $1\leq k \leq N-1,$ and
\begin{align}
\begin{split}
Y''_{1,j} & \in  \{Z_{1,j-1}''Y_j, -Y_{1,j-1}''X_j\},\\
Z''_{1,j} & \in  \{X_{1,j-1}''I_j, I_{1,j-1}''Z_j\},\\
I''_{1,j} & \in  \{X_{1,j-1}''Z_j, I_{1,j-1}''I_{j}\},
\end{split}
\end{align}
with $X''_{1,j}= -i Y''_{1,j}Z''_{1,j} \in\{Z_{1,j-1}''X_j,
Y_{1,j-1}''Y_j\}, $          and $2\leq j \leq N,$ and
\begin{align}
\begin{split}
&X'_{N,N}=X_N, ~Y'_{N,N}=Y_N, Z'_{N,N}=Z_N,\\
& X''_{1,1}=X_1,~~~~Y''_{1,1}=Y_1,~~~~ Z''_{1,1}=Z_1,\\
& I'_{N,N}=I_N,~~~~
 I''_{1,1}=I_1.
\end{split}
\end{align}

\begin{figure*}[htb]
\centering
\includegraphics[width=0.7 \textwidth]{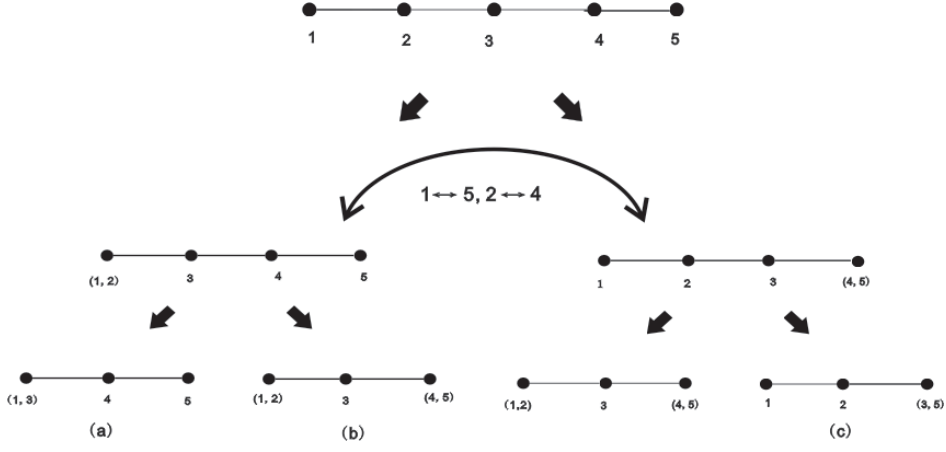}
\caption{\label{fig:5 qubits} Diagram of the three ways to derive
the GHZ argument for $|\phi_5\rangle$. Note that (a) and (c)
 can be transformed  to
each other   by simultaneously permuting qubit 1 and 5, qubit 2 and
4 based on the basic symmetry of $|\phi_5\rangle$.}
\end{figure*}

Now, we are in position to present our construction of the GHZ
argument  for $|\phi_5\rangle$.  Obviously the five-qubit cluster
state $|\phi_5\rangle$ may be rewritten either of the form
\begin{equation} \begin{split}
|\phi_5\rangle = & \frac{1}{2} \Big (
|+\rangle_{1,2}''|0\rangle_3|+\rangle_4|0\rangle_5
+|+\rangle_{1,2}''|0\rangle_3|-\rangle_4|1\rangle_5\\
&~~ + |-\rangle_{1,2}''|1\rangle_3|-\rangle_4|0\rangle_5
+|-\rangle_{1,2}''|1\rangle_3|+\rangle_4|1\rangle_5 \Big ),
\end{split}
\end{equation}
or
\begin{equation} \begin{split}
|\phi_5\rangle =& \frac{1}{2} \Big (
|+\rangle_1|0\rangle_2|+\rangle_3|0\rangle_{4,5}'
+|+\rangle_1|0\rangle_2|-\rangle_3|1\rangle_{4,5}'\\
&~~ + |-\rangle_1|1\rangle_2|-\rangle_3|0\rangle_{4,5}'
+|-\rangle_1|1\rangle_2|+\rangle_3|1\rangle_{4,5}' \Big ),
\end{split}
\end{equation}
by  regarding   the pair qubits 1, 2 and 4, 5  as two single objects
respectively. Fig.~\ref{fig:5 qubits} shows  three distinct ways to
derive the GHZ argument for $|\phi_5\rangle$. Note that   Eq.~(48)
and Eq.~(49) can be transformed to each other  by symmetrical
actions on the qubits subscripts: $1\longleftrightarrow 5$,
$2\longleftrightarrow 4$, thus we restrict our consideration to the
case of the form Eq.~(48).

Therefore, the GHZ argument for $|\phi_5\rangle$   expressed in
terms of Pauli-like operators is given by
\begin{align}
\begin{split}
&~~~~~~~~\{Z_{1,3}''X_4Z_5, Y_{1,3}''Y_4Z_{5},Y_{1,3}''X_4Y_{5},
Z_{1,3}''Y_4Y_{5}\},\\
&\Longrightarrow ~ \text{16 forms, \hspace{3.4cm}[see Fig.~\ref{fig:5 qubits}(a)]}\\
&~~~~~~~~\{Z_{1,2}''X_3Z'_{4,5},
Y_{1,2}''Y_3Z'_{4,5},Y_{1,2}''X_3Y'_{4,5},
Z_{1,2}''Y_3Y'_{4,5}\},\\
&~\Longrightarrow  \text{16 forms,\hspace{3.5cm}[see Fig.~\ref{fig:5 qubits}(b)]}\\
&~~~~~~~~ 1\longleftrightarrow 5,  2\longleftrightarrow 4,\\
&~\Longrightarrow  \text{16 forms, \hspace{3.3cm}  [see
Fig.~\ref{fig:5 qubits}(c)]}
\end{split}
\end{align}
where $Z_{1,3}'', Y_{1,3}'', Z_{1,2}'', Y_{1,2}'', Z_{4,5}',
Y_{4,5}'$ are depicted in Table~\ref{tab:13}. Hence, there are forty
eight distinct forms of the Greenberger-Horne-Zeilinger (GHZ)
argument for the five-qubit cluster state $|\phi_5\rangle.$

\begin{table}
\caption{\label{tab:13} $Z_{1,3}'', Y_{1,3}'', Z_{1,2}'', Y_{1,2}'',
Z_{4,5}', Y_{4,5}'.$ }
\begin{ruledtabular}
\begin{tabular}{cccccc}
$Z_{1,3}''$ & $Y_{1,3}''$ & $Z_{1,2}''$ & $Y_{1,2}''$  &$Z_{4,5}'$ & $Y_{4,5}'$ \\
\hline
   $Z_1X_2I_3$    &  $X_1I_2Y_3$     &      $X_1I_2$     & $Z_1Y_2$ & $I_4X_5$ & $-X_4Y_5$   \\
       $Y_1Y_2I_3$   &  $I_1Z_2Y_3$        &      $I_1Z_2$  & $-Y_1X_2$ & $Z_4I_5$ & $Y_4Z_5$   \\
       $X_1Z_2Z_3$     &  $-Z_1Y_2X_3$       &    &  &  &    \\
       $I_1I_2Z_3$          &  $Y_1X_2X_3$ &  & &  &\\
       \end{tabular}
\end{ruledtabular}
\end{table}

\section{ General GHZ theorem for the N-qubit cluster state ($N \geq 4$)}
\label{sec:N qubits}

Generally speaking, our method used in the previous section can be
easily extended to the case of the N-qubit system, since each form
of the GHZ argument for the N-qubit system of the one-dimensional
lattice can be eventually reduced to the three-qubit case in
Sec.~\ref{sec:3 qubits}.

More precisely,  the GHZ argument for $|\phi_N\rangle$ can be
expressed in terms of Pauli-like operators as
\begin{align}
\begin{split}
&C_j=\{Z_{1,j}''X_{j+1}Z_{j+2,N}', Y_{1,j}''Y_{j+1}Z_{j+2,N}',
Y_{1,j}''X_{j+1}Y_{j+2,N}',\\& \hspace{1cm}
Z_{1,j}''Y_{j+1}Y_{j+2,N}'\},\\
  &   \hspace{0.9cm} ~       \big( j=[N/2], [N/2]+1, \cdots, N-2 \big ),
\end{split}
\end{align}
and its permutations $C_j'$ as \begin{align} C_j' \myArrow[k=1, 2,
\cdots, {[N/2]}]{S_k} C_j,
\end{align}
where  $[N/2]$ denotes the largest integer $l$ with $l\leq N/2$, and
$S_k$ denotes the permuting action: changing the labeling of the
qubits subscripts with
\begin{align}
 k\longleftrightarrow(N+1-k).
\end{align}

Now we utilize the fact that $\mathcal {A}_{j+1}
 =\mathcal {A}_{j}\bigcup (E_j \mathcal {A}_{j})$,  by mathematics induction
 we arrive that
\begin{align}
\begin{split}
&Y_{1,j}''\in Z_{j-1}Y_j \mathcal {A}_j,\\
&Z_{1,j}''\in Z_j \mathcal {A}_j,\\
&I_{1,j}''\in  \mathcal {A}_j,\\
&X_{1,j}''=-iY_{1,j}''Z_{1,j}''\\
&\hspace{0.7cm}\in Z_{j-1}X_j \mathcal {A}_j,
\end{split}
\end{align}
where the operator set $\mathcal {A}_j$ is defined as
\begin{align}
\begin{split}
&\mathcal {A}_j=\{a_k, k=1, 2,\cdots, 2^{j-1}\}  \\
&~~~~\text{with}~~  a_k=\prod_{i\in T_k}E_i,
\end{split}
 \end{align}
and $T_k$ denotes a subset of the vertices set $\{1,2,\cdots,
j-1\}$.

Since the definition of $Y_{j+2,N}'$ etc.  can be transformed to
that of $Y_{1,N-j-1}''$ by symmetrical actions $S_k$ on the qubits
subscripts set $\{ 1,2,\cdots, N-j-1\}$, it turns out that
\begin{align}
\begin{split}
&Y_{j+2,N}'\in Y_{j+2}Z_{j+3} \mathcal {A}_{N-j-1}',\\
&Z_{j+2,N }'\in Z_{j+2} \mathcal {A}_{N-j-1}',\\
&I_{j+2,N}'\in  \mathcal {A}_{N-j-1}',\\
&X_{j+2,N}'=-iY_{j+2,N}'Z_{j+2,N }'\\
&\hspace{1.2cm} \in X_{j+2}Z_{j+3} \mathcal {A}_{N-j-1}',
\end{split}
\end{align}
with \begin{align} \mathcal {A}_{N-j-1}' \myArrow[k=1, 2, \cdots,
{N-j-1}]{S_k} \mathcal {A}_{N-j-1}.
\end{align}
Those equations allow us to construct the GHZ argument for the
cluster state of N-qubit in detail.

\section{ Bell inequalities}
\label{sec:Bell}

As is  demonstrated in Sec.~\ref{sec:N qubits}, we have presented a
complete construction of the GHZ argument for the cluster states of
N-qubit. In the following we extend the last considerations
concerning the associated Bell inequalities to the case of N-qubit
from the same conditions.

At the beginning let us define the Bell operator  in a standard
form:
\begin{align}
\mathcal {B}_{\phi_N}=&Z_{1,j}''X_{j+1}Z_{j+2,N}'+
Y_{1,j}''Y_{j+1}Z_{j+2,N}'\nonumber\\&
 -Y_{1,j}''X_{j+1}Y_{j+2,N}'  +
Z_{1,j}''Y_{j+1}Y_{j+2,N} \nonumber\\
=&(1+E_1')E_2'(1+E_3'),
\end{align}
with
\begin{align}
\begin{split}
E_1'&=X_{1,j}''Z_{j+1},\hspace{2cm}\\
E_2'&=Z_{1,j}''X_{j+1}Z_{j+2,N}',\\
E_3'&=Z_{j+1}X_{j+2,N}'.
\end{split}
\end{align}
It is easy to check that
\begin{align*}
&[E_m',E_n']=0,\hspace{0.3cm}    (E_n')^2=I, ~~~ \forall n, m=1, 2,
3.
\end{align*}
A simple computation yields that
\begin{align}
\mathcal{B}_{\phi_N}^2=4(1+E_1')(1+E_3').
\end{align}
It is concluded that $\norm{ \mathcal{B}_{\phi_N}^2} =16$ and $
\norm{\mathcal{B}_{\phi_N} }=4.$ This shows that ${B}_{\phi_N}$
reaches 4 when evaluated on ${\phi_N}$ (${\phi_N}$ is the common
eigenstate of $\mathcal{B}_{\phi_N}$).

In a local realistic theory, the correlation function of the
measurements performed by the observers is the average of the
outcomes over many runs of the experiment. The classical correlation
functions corresponding to $\mathcal{B}_{\phi_N}$  is
\begin{align}
|&\langle z_{1,j}''x_{j+1}z_{j+2,N}'\rangle+ \langle
y_{1,j}''y_{j+1}z_{j+2,N}'\rangle -\langle
y_{1,j}''x_{j+1}y_{j+2,N}' \rangle  \nonumber \\
& +\langle z_{1,j}''y_{j+1}y_{j+2,N}'\rangle| \leq2,
\end{align}
by grouping 1 and 2 together in an analogous way to that presented
in \cite{Mermin, ABD}.

In Table~\ref{tab:456} we give out   all the  Bell operators in the
form of $\mathcal {B}_{\phi_N}=(1+E_1')E_2'(1+E_3')$ with $N=4,5,6$
which in turn introduce Bell inequalities. Our result generalizes
those investigations in \cite{SASA}, and is an extended version of
\cite{GCR}. Moreover, our method can be easily extended to N qubits.

From Eq.~(54) and Eq.~(56), one can immediately infer that
\begin{align}
\begin{split}
&E_1'\in E_j \mathcal {A}_j,\\
&E_2'\in \mathscr{A}=E_{j+1}\mathcal {A}_j * \mathcal {A}_{N-j-1}',\\
 &E_3'\in E_{j+2} \mathcal {A}_{N-j-1}',
 \end{split}
\end{align}
where the notion $*$ is defined as  $A*B=\{xy,x\in A, y\in B\}$.

In Sec.~\ref{sec:4 and 5 qubits}, it is pointed out that these Bell
operators are all maximally violated by the cluster state, but not
only by the cluster state, since they  are  not a GHZ-Mermin
experiment at whose some site there is only one measurement. For
$|\phi_N\rangle$ combining $2^{N-3}$ of those  Bell operators, we
could obtain a new Bell operator such that only $|\phi_N\rangle$
violate it maximally.
 The N-qubit Bell operator is then defined  as
\begin{align}
\begin{split}
 \mathscr{B}_{\phi_N}=&(1+E_j)\sum_{ \omega\in \mathscr{A} }\omega
(1+E_{j+2})\\
=& E_{j+1}\prod^N_{\begin{subarray}{c} m=1,\\
 m\neq j+1
\end{subarray}}(1+E_m).
\end{split}
\end{align}

To give a simple example,
\begin{align}
\mathscr{B}_{|\phi_4\rangle} =&(1+E_1)E_2(1+E_3)(1+E_4),\nonumber \\
&\hspace{1cm}(E_2\longleftrightarrow E_3), \\
\mathscr{B}_{|\phi_5\rangle}=& (1+E_1)(1+E_2)E_3(1+E_4)(1+E_5),\nonumber\\
&\hspace{1cm}(E_3\longleftrightarrow E_4, E_2),\\
\mathscr{B}_{|\phi_6\rangle}=& (1+E_1)(1+E_2)(1+E_3)E_4(1+E_5)(1+E_6),\nonumber\\
&\hspace{1cm}(E_4\longleftrightarrow E_3, E_2,E_5).
\end{align}

\begin{table}
\caption{\label{tab:456}The Bell operators for $ |\phi_N\rangle$
 $\mathcal {B}_{\phi_N}=(1+E_1')E_2'(1+E_3')$, with  N=4,
5, 6.}
\begin{ruledtabular}
\begin{tabular}{c|cccc|c}
$ |\phi_N\rangle$ &  & $E_1'$   & $E_2'$& $E_3'$& \\
\hline
 $N=4$   & $(1,2),3,4$  & $E_2$  & $E_3$
&$E_4$& $1\longleftrightarrow 4$  \\
& & $E_1E_2$ & $E_1E_3$ & & $2\longleftrightarrow 3$ \\ \cline{1-6}

$N=5$ & $(1,2),3,(4,5)$  & $E_2$  & $E_3$ &$E_4$&\\
 &   & $E_1E_2$  & $E_1E_3$ &$E_4E_5$& \\
 &  &   & $E_3E_5$ & & \\
&  &   & $E_1E_3E_5$ & &
\\  \cline{2-6}
 & $(1,3),4,5$  & $E_3$  & $E_4$ &$E_5$& $1\longleftrightarrow 5$\\
&   & $E_1E_3$  & $E_1E_4$ && $2\longleftrightarrow 4$\\
&  & $E_2E_3$  & $E_2E_4$ &&\\
&  & $E_1E_2E_3$  & $E_1E_2E_4$ &&\\
\hline
$N=6$ & $(1,3),4,(5,6)$  & $E_3$  & $E_4$ &$E_5$&$1\longleftrightarrow 6$\\
&  & $E_1E_3$  &$E_1E_4$  &$E_5E_6$&$2\longleftrightarrow 5$\\
&  & $E_2E_3$  &$E_2E_4$  &&$3\longleftrightarrow 4$\\
&  & $E_1E_2E_3$  &$E_4E_6$  &&\\
&  &   &$E_1E_2E_4$  &&\\
&  &   &$E_1E_4E_6$  &&\\
&  &   &$E_2E_4E_6$  &&\\
&  &   &$E_1E_2E_4E_6$  &&\\
\cline{2-6} & (1,4),5,6 & $E_4$  & $E_5$ & $E_6$ &
$1\longleftrightarrow 6$\\
&  & $E_1E_4$  & $E_1E_5$ &  &
$2\longleftrightarrow 5$\\
&  & $E_2E_4$  & $E_2E_5$ &  &
$3\longleftrightarrow 4$\\
&  & $E_3E_4$  & $E_3E_5$ &  &
\\
&  & $E_1E_2E_4$  & $E_1E_2E_5$ &  &\\
&  & $E_1E_3E_4$  & $E_1E_3E_5$ &  &\\
&  & $E_2E_3E_4$  & $E_2E_3E_5$ &  &\\
&  & $E_1E_2E_3E_4$  & $E_1E_2E_3E_5$ &  &\\
\end{tabular}
\end{ruledtabular}
\end{table}

\section{ conclusion}
\label{sec:summary}

In conclusion, we have introduced a general method to derive the GHZ
argument for the cluster states of N-qubit in the case of the
one-dimensional lattice.  We show that there are eight distinct
forms of the GHZ argument for $|\phi_4\rangle$ and forty eight
distinct forms for $|\phi_5\rangle$ respectively. The proof is
obtained by regarding the pair qubits as a single object and
constructing the new Pauli-like operators. Our method can be easily
extended to the case of the N-qubit system. In addition, we discuss
the associated Bell inequalities and  therefore obtain a new Bell
operator such that only $|\phi_N\rangle$ violate it maximally.

\acknowledgments

The authors would like to thank Chuang Ye Liu for useful discussions
and suggestions. This work is supported by the NNSF of China, the
CAS, the National Fundamental Research Program (Grant No.
2006CB921900), and partially supported by the National Science
Foundation of China  under Grant 10775175.

%

\end{document}